# Illusion optics: The optical transformation of an object into another object

Yun Lai,* Jack Ng,* HuanYang Chen, DeZhuan Han, JunJun Xiao, Zhao-Qing Zhang† and C. T. Chan†

Department of Physics

The Hong Kong University of Science and Technology

Clear Water Bay, Kowloon, Hong Kong, China

## Abstract

We propose to use transformation optics to generate a general illusion such that an arbitrary object appears to be like some other object of our choice. This is achieved by using a remote device that transforms the scattered light outside a virtual boundary into that of the object chosen for the illusion, regardless of the profile of the incident wave. This type of illusion device also enables people to see through walls. Our work extends the concept of cloaking as a special form of illusion to the wider realm of illusion optics.

* These authors contributed equally to this work.
† E-mail: To whom correspondence should be addressed. E-mail: phzzhang@ust.hk (Z. Q. Zhang); phchan@ust.hk (C. T. Chan)

As the saying goes, "seeing is believing." Throughout history, witnessing with the eyes has been used as proof of existence or as evidence. On the other hand, the effects of illusions, such as mirages, have been well known to lead people to draw incorrect conclusions, sometimes with dire consequences. Recently, the rapid development of transformation optics [1-22] has enabled the design of new materials that can steer light along arbitrary curves and the implementation is made possible by a new kind of man-made materials called metamaterials [23-27]. Among various novel applications, the most fascinating is a cloaking device designed to bend light around a concealed region, rendering any object inside the region "invisible" [1-10]. Cloaking can be regarded as creating an illusion of free space. In this paper, we discuss a more generalized concept of illusion: making an object of arbitrary shape and material properties appear exactly like another object of some other shape and material makeup. Using transformation optics, we design an illusion device consisting of two distinct pieces of metamaterials, which are called the "complementary medium" and the "restoring medium". The complementary medium concept, which was first proposed by Pendry *et al.* to make focusing lenses [28, 29], is applied here to "cancel" a piece of space optically, including the object [21, 22]. Then, the restoring medium restores the cancelled space with a piece of the illusion space that is embedded with the other object chosen for the illusion. Regardless of the profile and the direction of the incident light, the illusion device can transform the scattered light outside a virtual boundary into that of the second (illusion) object; it therefore creates a stereoscopic illusion for any observer outside the virtual boundary.

The principle behind this illusion device is not light bending, but rather the exact cancellation and restoration of the optical path of light within the virtual boundary. Unlike previous light-bending cloaking devices [1-10], the constitutive parameters of the illusion device do not have a complex spatial distribution or any singularities. More surprisingly, the illusion device works at a distance from the object. An interesting implication of this "remote" feature is the ability to open a virtual aperture in a wall so that one can peep through walls in a noninvasive manner. By making an illusion of a "hole" in a wall, one can see through the wall as if the wall has actually had a hole, and for this purpose, monochromic functionality is sufficient.

A simple schematic diagram illustrating our idea is shown in Fig. 1. In Fig. 1(a), an illusion device is placed next to a domain that contains a man (the object). The passive device causes any observer outside the virtual boundary (the dashed curves) to see the image of a woman (the illusion) inside the illusion space depicted in Fig. 1(b). We will show that we can design such an illusion device, which makes the electromagnetic fields outside the virtual boundary in both the real and illusion spaces exactly the same, irrespective of the profile of the incident waves. A blueprint for the device is shown in Fig. 1(c), in which there are two regions. Region 2 includes the "complementary medium" used to annihilate the optical signature of the man and region 1 includes the "restoring medium" that creates the image of the woman. Both media are designed using transformation optics [1-4]. The complementary medium is formed by a coordinate transformation of folding region 3, which contains the man, into region 2. The restoring medium is formed by a coordinate transformation of compressing region 4 in Fig. 1(d), which contains the illusion, into region 1. The permittivity and permeability tensors of both media in the illusion device can be expressed as: $\boldsymbol{\varepsilon}^{(2)} = \mathbf{A}\boldsymbol{\varepsilon}^{(3)}\mathbf{A}^{\mathbf{T}} / \det \mathbf{A}$, $\boldsymbol{\mu}^{(2)} = \mathbf{A}\boldsymbol{\mu}^{(3)}\mathbf{A}^{\mathbf{T}} / \det \mathbf{A}$, $\boldsymbol{\varepsilon}^{(1)} = \mathbf{B}\boldsymbol{\varepsilon}^{(4)}\mathbf{B}^{\mathbf{T}} / \det \mathbf{B}$ and $\boldsymbol{\mu}^{(1)} = \mathbf{B}\boldsymbol{\mu}^{(4)}\mathbf{B}^{\mathbf{T}} / \det \mathbf{B}$, where $\boldsymbol{\varepsilon}^{(i)}$ and $\boldsymbol{\mu}^{(i)}$ are the permittivity and permeability tensors in region $i$, $\mathbf{A}$ and $\mathbf{B}$ are the Jacobian transformation tensors with components $A_{ij} = \partial x_i^{(2)} / \partial x_j^{(3)}$ and $B_{ij} = \partial x_i^{(1)} / \partial x_j^{(4)}$, corresponding to the coordinate transformations of folding region 3 into region 2 and compressing region 4 into region 1, respectively.

The electromagnetic fields in the complementary and the restoring media can also be obtained from transformation optics [1-4] as : $\mathbf{E}^{(2)} = (\mathbf{A}^{\mathbf{T}})^{-1}\mathbf{E}^{(3)}$, $\mathbf{H}^{(2)} = (\mathbf{A}^{\mathbf{T}})^{-1}\mathbf{H}^{(3)}$, $\mathbf{E}^{(1)} = (\mathbf{B}^{\mathbf{T}})^{-1}\mathbf{E}^{(4)}$ and $\mathbf{H}^{(1)} = (\mathbf{B}^{\mathbf{T}})^{-1}\mathbf{H}^{(4)}$, where $\mathbf{E}^{(i)}$ and $\mathbf{H}^{(i)}$ are the electric and magnetic fields in region $i$, respectively. From the matching of the boundary conditions on surface $a$ (the red solid curve) between the complementary medium and the restoring medium, we have $\mathbf{E}_t^{(2)}(a) = \mathbf{E}_t^{(1)}(a)$ and $\mathbf{H}_t^{(2)}(a) = \mathbf{H}_t^{(1)}(a)$, where subscript $t$ indicates transverse components along the surface. Both the folding transformation, $\mathbf{A}$, and compression transformations, $\mathbf{B}$, map one part of the virtual boundary, i.e. surface $c$ (the red dashed curves), to surface $a$. If this one-to-one mapping from $c$ to $a$ is the same for both $\mathbf{A}$ and **B**, then we can obtain from transformation optics that $\mathbf{E}_t^{(3)}(c) = \mathbf{E}_t^{(4)}(c)$ and $\mathbf{H}_t^{(3)}(c) = \mathbf{H}_t^{(4)}(c)$ on surface $c$. In addition, we also have $\mathbf{E}_t^{(1)}(d) = \mathbf{E}_t^{(4)}(d)$ and $\mathbf{H}_t^{(1)}(d) = \mathbf{H}_t^{(4)}(d)$ on the other part of the virtual boundary, i.e., surface $d$ (the blue dashed curves), as long as $d$ is not changed during transformation **B**. Therefore, the tangential components of the electromagnetic fields on the whole virtual boundary (including $c$ and $d$) are exactly the same in the real and illusion spaces, and, consequently, by the uniqueness theorem, the electromagnetic fields outside are also exactly the same. Any observer outside the virtual boundary will see electromagnetic waves as if they were scattered from the illusion object (the woman and nothing else), and thus an illusion is created. A detailed proof is provided in the Auxiliary Material [30].

In the following, we describe full wave simulations using a finite element solver (Comsol Multiphysics) to demonstrate the explicit effect of an illusion device that transforms a dielectric spoon of $\varepsilon_o = 2$ into a metallic cup of $\varepsilon_i = -1$ in two dimensions. The electromagnetic waves can be decoupled into TE waves ($E$ along the $z$ direction) and TM waves ($H$ along the $z$ direction); we show only the TE results for brevity (the parameters can be tuned to work for both TE and TM waves). Figs. 2(a) and 2(c) plot, respectively, the scattering patterns of the dielectric spoon and the metallic cup, under the illumination of a TE plane wave (propagating from left to right) of wavelength $\lambda = 0.25$ unit. In Fig. 2(b), an illusion device is placed beside the spoon. The scattering pattern around the spoon and the illusion device is altered in such a way that it appears as if there is only a metallic cup. This can be clearly seen by comparing the field patterns of the spoon plus the illusion device shown in Fig. 2(b) with that of the metallic cup shown in Fig. 2(c). The field patterns are indeed identical outside the virtual boundary. Inside the virtual boundary, the field patterns in Figs. 2(b) and 2(c) are completely different. The fields between the spoon and the illusion device are strong due to the excitation of surface resonances induced by the multiple scattering of light between the spoon and the illusion device. We note that the illusion effect is a steady state phenomenon that takes some time to establish. More simulation results under different kinds of incident waves can be found in the Auxiliary Material [30].

The illusion device in Fig. 2(b) is composed of four parts. The lower trapezoidal part is the "complementary medium" formed by a simple coordinate transformation of $y^{(2)} = -y^{(3)}/2$. It is composed of a negative index homogeneous medium of $\varepsilon_z^{(2)} = -2$, $\mu_x^{(2)} = -2$ and $\mu_y^{(2)} = -0.5$, with an embedded "anti-object" of the dielectric spoon with $\varepsilon_{oz}^{(2)} = -4$ and $\boldsymbol{\mu}_o^{(2)} = \boldsymbol{\mu}^{(2)}$. The upper left triangular part, the upper right triangular part, and the upper middle rectangular part collectively constitute the "restoring medium". The upper left and right triangular parts are composed of an homogeneous medium with $\varepsilon_z^{(1)} = 4$, $\mu_{xx}^{(1)} = 4$, $\mu_{yy}^{(1)} = 20.5$ and $\mu_{xy}^{(1)} = \pm 9$, formed by the coordinate transformations of $y^{(1)} \mp 3(x \pm 0.6) = 1/4 \cdot \left( y^{(4)} \mp 3(x \pm 0.6) \right)$, respectively. The upper middle rectangular part is composed of an homogeneous medium of $\varepsilon_z^{(1)} = 4$, $\mu_x^{(1)} = 4$ and $\mu_y^{(1)} = 0.25$, with an embedded compressed version of the metallic cup illusion of $\varepsilon_{iz}^{(1)} = -4$ and $\boldsymbol{\mu}_i^{(1)} = \boldsymbol{\mu}^{(1)}$, formed by the coordinate transformation of $y^{(1)} - 0.6 = 1/4 \cdot \left( y^{(4)} - 0.6 \right)$. It is important to note that the permittivity and permeability of the illusion device are both composed of simple homogeneous media and this simplicity is a consequence of the simple coordinate transformations applied here. They do not bend straight light paths into curved ones as in light-bending cloaking devices [1-10].

The complementary medium is obtained from the transformation optics of folded geometry (see, for example, Leonhardt *et al.* [10]). It is composed of left-handed metamaterials with simultaneously negative permittivity and permeability. The medium can be isotropic if we apply a transformation of $y^{(2)} = -y^{(3)}$ instead of $y^{(2)} = -y^{(3)}/2$. This kind of metamaterial has been extensively studied in the application of the superlens [28], and it has been fabricated by various resonant structures at various frequencies [23-27]. The other key component of the illusion device is the restoring medium, which projects the optical illusion of the metallic cup. It is composed of the homogeneous medium with positive but anisotropic permeability. This kind of medium can be designed from layer-structured metamaterials [15].

We note that some special illusion tricks by image projection using transformation optics have been discovered, such as the shifted-position image of an object inside a metamaterial shell [16], the cylindrical superlens [17], the "superscatterer" [18], the "reshaper" [19] and the "super absorber" [20]. Recently, we proposed an approach to realize "cloaking at a distance" by using an "anti-object" [21, 22]. Here, by combining the "anti-object" cloaking functionality and the image projection functionality, we achieve a general form of illusion optics such that an object can be disguised into something else and the illusion device itself is invisible. This general form of illusion optics with arbitrary shape and generalized topology is proved mathematically as it is designed with transformation optics and the functionality is also demonstrated numerically. From a multiple scattering point of view, the illusion optics is in fact rather remarkable as it is by no means obvious that the anti-object cancelling and the image projection functionality do not obstruct or interfere with each other.

Another interesting application of our illusion device is that it enables people to open a virtual hole in a wall or obstacle. As our illusion device works at a distance from the object, it is capable of transforming only one part of an object into an illusion of free space, thus rendering that part invisible while leaving the rest of the object unaffected. By making one part of the wall invisible (i.e., making an illusion of a "hole"), we can then see through the wall and obtain information from the other side. In Fig. 3(a), we see that a wall of $\varepsilon_o = -1$ with a width of 0.2 units is capable of blocking most of the energy of the TE electromagnetic waves radiating from a point source of $\lambda = 0.25$ unit placed at $(-0.7, 0)$. When the illusion device is placed on the right side of the wall, as shown in Fig. 3(b), the electromagnetic waves can penetrate through the wall as well as the illusion device and arrive on the right side. This effect can also be understood as the tunneling of electromagnetic waves via the high-intensity surface waves localized at the interface between the wall and the complementary medium. The phase information is accurately corrected by the restoring medium in the illusion device, such that the transmitted field patterns on the right side become the same as those of the electromagnetic waves penetrating through a real hole, as shown in Fig. 3(c). Thus, an observer on the right side can peep through the virtual hole as if he/she is peeping through a real hole at the working frequency of the illusion device. The constitutive parts of the illusion device are similar to that in Fig. 2(b) and described in detail in the Auxiliary Material [30]. Similarly, an object hidden in a container can be completely revealed by using the illusion optics to change the container into an illusion of free space. This is also demonstrated in the Auxiliary Material [30].

In principle, the illusion optics allows us to remotely change the optical response of an object into that of any other object chosen for illusion at a selected frequency, without the need to change the constituents and shape of the true object or even cover its surface. This opens up interesting possibilities. For instance, an illusion waveguide or photonic crystal would allow the control of light propagation in actual free space; an illusion tip might perform near-field scanning optical microscopy without physically approaching a surface. However, the theoretical foundation of the illusion device is transformation optics and, as such, our theory relies on the validity and accuracy of a linear continuous medium that describes the homogenized electromagnetic fields in metamaterials. This requirement is crucial in the interface between the complementary medium and the "cancelled" object due to the high-intensity local fields as well as rapid oscillations there. The range of the virtual boundary also plays an important role. When it is large, the field at the boundary will be large as well. Another issue that we have not considered is loss, which will degrade the illusion effect unless the object is close to the device. If these issues and challenges can be solved with advances in metamaterial technologies, we should be able to harness the power of transformation optics to create illusions.

This work was supported by Hong Kong Central Allocation Grant No. HKUST3/06C. Computation resources are supported by Shun Hing Education and Charity Fund. We thank Dr. KinHung Fung, ZhiHong Hang, Jeffrey ChiWai Lee and HuiHuo Zheng for helpful discussions.

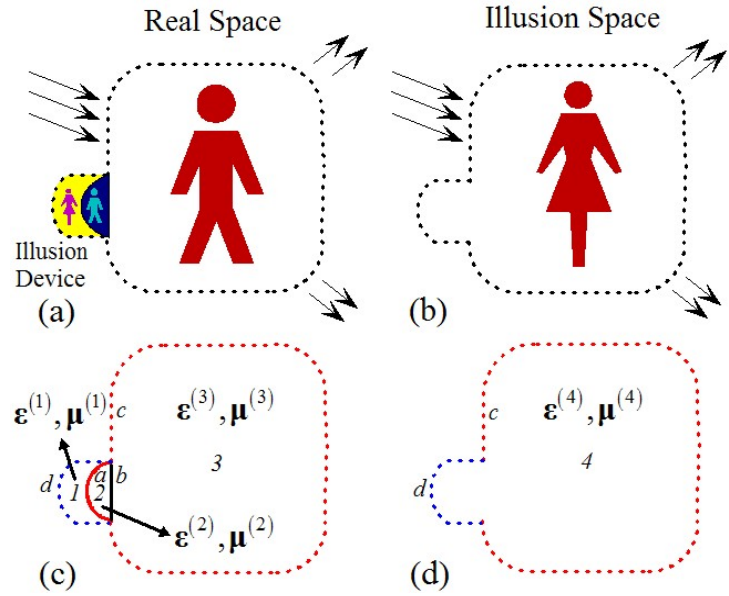

Fig. 1 (color online). The working principle of an illusion device that transforms the stereoscopic image of the object (a man) into that of the illusion (a woman). (a) The man (the object) and the illusion device in real space. (b) The woman (the illusion) in the illusion space. (c) The physical description of the system in real space. The illusion device is composed of two parts, the complementary medium (region 2) that optically "cancels" a piece of space including the man (region 3), and the restoring medium (region 1) that restores a piece of the illusion space including the illusion (region 4 in (d)). Both real and illusion spaces share the same virtual boundary (dashed curves).

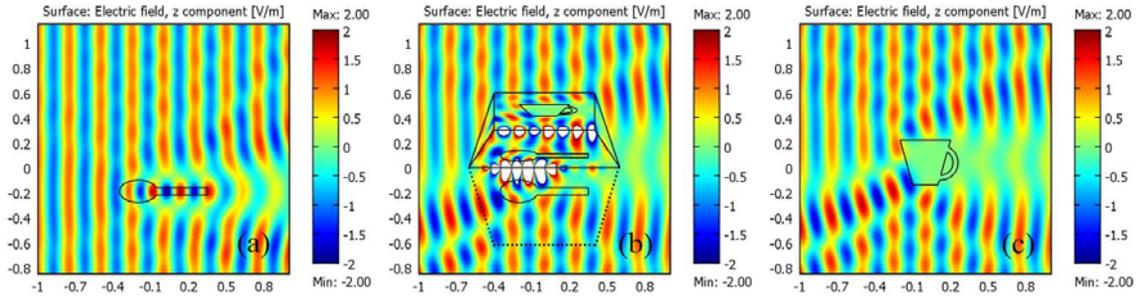

Fig. 2 (color online). A numerical demonstration of transforming the stereoscopic image of a dielectric spoon of $\varepsilon_o = 2$ (the object) into that of a metallic cup of $\varepsilon_i = -1$ (the illusion) through an illusion device, under an incident TE plane wave from the left. (a) The scattering pattern of the dielectric spoon. (b) The scattering pattern of the dielectric spoon is changed by the illusion device. Outside the virtual boundary, the scattering pattern becomes the same as that of the metallic cup, which is shown in (c).

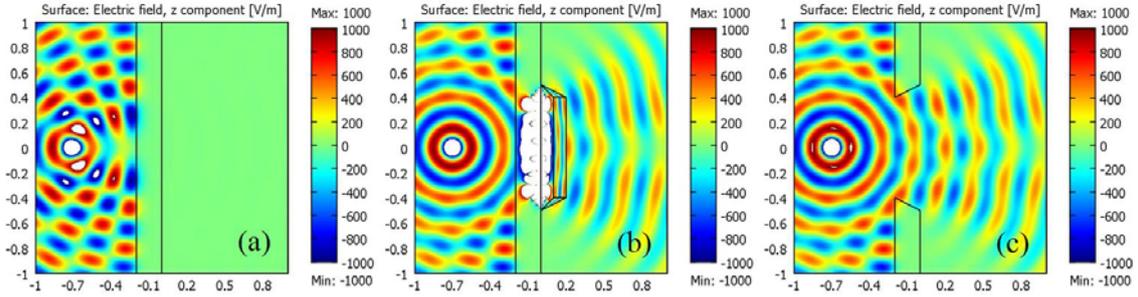

Fig. 3 (color online). The illusion device can create the illusion of a hole so that people can see through a wall at a selected frequency. (a) The electromagnetic radiation from a TE point source on the left side is blocked by a slab of $\varepsilon_o = -1$. (b) When an illusion device is attached to the wall, the electromagnetic radiation can now tunnel through the wall to the right side. The far field radiation pattern is exactly the same as that of the radiation through a real hole, which is shown in (c).

**Auxiliary material of "Illusion optics: The optical transformation of an object into another object"**

Yun Lai*, Jack Ng*, Huanyang Chen, DeZhuan Han, JunJun Xiao, Zhao-Qing Zhang[†] and C. T. Chan[†]

*Department of Physics*
*The Hong Kong University of Science and Technology*
*Clear Water Bay, Kowloon, Hong Kong, China*

Part A: A rigorous proof of the illusion optics in 3D by transformation optics

We shall prove here that by using the complementary medium and the restoring medium designed from transformation optics, we are able to transform an object into an illusion of another object of our choice. Both the object and the illusion can be anisotropic and/or inhomogeneous.

Consider the configuration depicted in Fig. A1. The real space is divided into four domains: regions 1, 2, 3, and the region outside surfaces *c* and *d*. The illusion space is divided into two domains: region 4 and the region outside surfaces *c* and *d*. Under light illumination, there will be a solution in each of these regions. Our task is to prove that under arbitrary light illumination, the solution outside surfaces *c* and *d* is the same for both the real space and the illusion space, such that any outside observer would think that he/she has seen the illusion while what are really there are the object and the illusion device.

We parameterize region *i* by the generalized curved coordinates $(u^{(i)}, v^{(i)}, w^{(i)})$, as depicted in Fig. A2. The permittivity and permeability tensors in region *i* are respectively denoted as $\boldsymbol{\varepsilon}^{(i)}(u^{(i)}, v^{(i)}, w^{(i)})$ and $\boldsymbol{\mu}^{(i)}(u^{(i)}, v^{(i)}, w^{(i)})$, and the electric and magnetic fields of region *i* are respectively denoted as $\mathbf{E}^{(i)}$ and $\mathbf{H}^{(i)}$. Region 3 is a piece of space embedded with the object that we want to transform into something else. Region 2 is composed of the complementary medium of region 3, whose dielectric properties are obtained by the coordinate transformation of folding region 3 into region 2:

$$\boldsymbol{\varepsilon}^{(2)} = \mathbf{A}\boldsymbol{\varepsilon}^{(3)}\mathbf{A}^{\mathrm{T}} / \det \mathbf{A},$$
$$\boldsymbol{\mu}^{(2)} = \mathbf{A}\boldsymbol{\mu}^{(3)}\mathbf{A}^{\mathrm{T}} / \det \mathbf{A}, \tag{1}$$

with each point on surface *c* being mapped to a point on surface *a* in a one-to-one and continuous manner, and each point on surface *b* being mapped back to itself. Here,

$$\mathbf{A} = \begin{bmatrix} \dfrac{\partial u^{(2)}}{\partial u^{(3)}} & \dfrac{\partial u^{(2)}}{\partial v^{(3)}} & \dfrac{\partial u^{(2)}}{\partial w^{(3)}} \\ \dfrac{\partial v^{(2)}}{\partial u^{(3)}} & \dfrac{\partial v^{(2)}}{\partial v^{(3)}} & \dfrac{\partial v^{(2)}}{\partial w^{(3)}} \\ \dfrac{\partial w^{(2)}}{\partial u^{(3)}} & \dfrac{\partial w^{(2)}}{\partial v^{(3)}} & \dfrac{\partial w^{(2)}}{\partial w^{(3)}} \end{bmatrix} \tag{2}$$

is the Jacobian transformation tensor of the folding transformation. From transformation optics, the electromagnetic fields of region 2 and 3 are related by

$$\mathbf{A}^{\mathrm{T}}\mathbf{E}^{(2)} = \mathbf{E}^{(3)},$$
$$\mathbf{A}^{\mathrm{T}}\mathbf{H}^{(2)} = \mathbf{H}^{(3)}. \tag{3}$$

It can be shown that the boundary conditions on surface *b* are fulfilled. By exploiting our freedom to select the parametric coordinate *w* such that surface *b* is a constant level surface of $w^{(2)}$ and $w^{(3)}$, we have on surface *b*:

$$\frac{\partial w^{(2)}}{\partial v^{(3)}} = \frac{\partial w^{(2)}}{\partial u^{(3)}} = 0. \tag{4}$$

Furthermore, since each point on surface *b* is being mapped back to itself, the parametric coordinate $\left(u^{(2)}, v^{(2)}\right)$ can be chosen to exactly coincide with $\left(u^{(3)}, v^{(3)}\right)$ on surface *b*, which gives on surface *b*

$$\frac{\partial u^{(2)}}{\partial v^{(3)}} = \frac{\partial v^{(2)}}{\partial u^{(3)}} = 0,$$
$$\frac{\partial v^{(2)}}{\partial v^{(3)}} = \frac{\partial u^{(2)}}{\partial u^{(3)}} = 1. \tag{5}$$

Substituting Eqs. (4) and (5) into Eq. (2), we obtain, on surface *b*,

$$\mathbf{A}(b \to b) = \begin{bmatrix} 1 & 0 & \dfrac{\partial u^{(2)}}{\partial w^{(3)}} \\ 0 & 1 & \dfrac{\partial v^{(2)}}{\partial w^{(3)}} \\ 0 & 0 & \dfrac{\partial w^{(2)}}{\partial w^{(3)}} \end{bmatrix}. \qquad (6)$$

Using Eqs. (3) and (6), it can be shown that the boundary conditions on surface $b$ are fulfilled:

$$\begin{aligned} E^{(3)}_{v^{(3)}} &= E^{(2)}_{v^{(2)}}, & E^{(3)}_{u^{(3)}} &= E^{(2)}_{u^{(2)}}, \\ H^{(3)}_{v^{(3)}} &= H^{(2)}_{v^{(2)}}, & H^{(3)}_{u^{(3)}} &= H^{(2)}_{u^{(2)}}. \end{aligned} \qquad (7)$$

We next consider the mapping of surface $c$ to surface $a$ in real space. On surface $a$ and on the side of region 2, we can again exploit our freedom to choose the parametric coordinate such that surface $a$ is a constant level surface of $w^{(2)}$, and similarly surface $c$ is a constant level surface of $w^{(3)}$, such that, on surface $a$,

$$\frac{\partial w^{(2)}}{\partial v^{(3)}} = \frac{\partial w^{(2)}}{\partial u^{(3)}} = 0. \qquad (8)$$

The transformation Jacobian is then

$$\mathbf{A}(c \to a) = \begin{bmatrix} \dfrac{\partial u^{(2)}}{\partial u^{(3)}} & \dfrac{\partial u^{(2)}}{\partial v^{(3)}} & \dfrac{\partial u^{(2)}}{\partial w^{(3)}} \\ \dfrac{\partial v^{(2)}}{\partial u^{(3)}} & \dfrac{\partial v^{(2)}}{\partial v^{(3)}} & \dfrac{\partial v^{(2)}}{\partial w^{(3)}} \\ 0 & 0 & \dfrac{\partial w^{(2)}}{\partial w^{(3)}} \end{bmatrix}. \qquad (9)$$

Using Eqs. (3) and (9), we obtain the relations of the tangential fields on surface $c$ in real space and surface $a$ as:

$$\begin{aligned} E^{(3)}_{u^{(3)}} &= \frac{\partial u^{(2)}}{\partial u^{(3)}} E^{(2)}_{u^{(2)}} + \frac{\partial v^{(2)}}{\partial u^{(3)}} E^{(2)}_{v^{(2)}}, \\ E^{(3)}_{v^{(3)}} &= \frac{\partial u^{(2)}}{\partial v^{(3)}} E^{(2)}_{u^{(2)}} + \frac{\partial v^{(2)}}{\partial v^{(3)}} E^{(2)}_{v^{(2)}}, \end{aligned} \qquad (10)$$

and the expressions of the magnetic fields are similar.

On the other hand, region 1 is composed of the restoring medium with solutions $\mathbf{E}^{(1)}$ and $\mathbf{H}^{(1)}$. Since $\mathbf{E}^{(1)}$ and $\mathbf{H}^{(1)}$ are the solutions in real space, they must satisfy the boundary condition on surface *a*. Accordingly, the transverse component of $\mathbf{E}^{(1)}$ and $\mathbf{H}^{(1)}$ equals that of $\mathbf{E}^{(2)}$ and $\mathbf{H}^{(2)}$ on surface *a*, respectively:

$$E_{v^{(1)}}^{(1)} = E_{v^{(2)}}^{(2)}, \quad E_{u^{(1)}}^{(1)} = E_{u^{(2)}}^{(2)}, \\ H_{v^{(1)}}^{(1)} = H_{v^{(2)}}^{(2)}, \quad H_{u^{(1)}}^{(1)} = H_{u^{(2)}}^{(2)}. \tag{11}$$

Substituting Eq. (11) into Eq. (10), we obtain

$$E_{u^{(3)}}^{(3)} = \frac{\partial u^{(2)}}{\partial u^{(3)}} E_{u^{(1)}}^{(1)} + \frac{\partial v^{(2)}}{\partial u^{(3)}} E_{v^{(1)}}^{(1)}, \\ E_{v^{(3)}}^{(3)} = \frac{\partial u^{(2)}}{\partial v^{(3)}} E_{u^{(1)}}^{(1)} + \frac{\partial v^{(2)}}{\partial v^{(3)}} E_{v^{(1)}}^{(1)}, \tag{12}$$

and the expressions of the magnetic fields are similar. We note that the dielectric properties of region 1 are determined by the coordinate transformation of compressing region 4 in the illusion space into region 1:

$$\boldsymbol{\varepsilon}^{(1)} = \mathbf{B} \boldsymbol{\varepsilon}^{(4)} \mathbf{B}^\mathbf{T} / \det \mathbf{B} \\ \boldsymbol{\mu}^{(1)} = \mathbf{B} \boldsymbol{\mu}^{(4)} \mathbf{B}^\mathbf{T} / \det \mathbf{B} \tag{13}$$

with each point on surface *c* being mapped to a point on surface *a* in a one-to-one and continuous manner, and each point on surface *d* being mapped back to itself. Here

$$\mathbf{B} = \begin{bmatrix} \dfrac{\partial u^{(1)}}{\partial u^{(4)}} & \dfrac{\partial u^{(1)}}{\partial v^{(4)}} & \dfrac{\partial u^{(1)}}{\partial w^{(4)}} \\ \dfrac{\partial v^{(1)}}{\partial u^{(4)}} & \dfrac{\partial v^{(1)}}{\partial v^{(4)}} & \dfrac{\partial v^{(1)}}{\partial w^{(4)}} \\ \dfrac{\partial w^{(1)}}{\partial u^{(4)}} & \dfrac{\partial w^{(1)}}{\partial v^{(4)}} & \dfrac{\partial w^{(1)}}{\partial w^{(4)}} \end{bmatrix} \tag{14}$$

is the Jacobian transformation tensor of the compressing transformation. The electromagnetic fields in the restoring medium can also be obtained from transformation optics:

$$\mathbf{B}^\mathbf{T} \mathbf{E}^{(1)} = \mathbf{E}^{(4)}, \\ \mathbf{B}^\mathbf{T} \mathbf{H}^{(1)} = \mathbf{H}^{(4)}. \tag{15}$$

On surface *a* and on the side of region 1, the transformation Jacobian is

$$\mathbf{B}(c \to a) = \begin{bmatrix} \dfrac{\partial u^{(1)}}{\partial u^{(4)}} & \dfrac{\partial u^{(1)}}{\partial v^{(4)}} & \dfrac{\partial u^{(1)}}{\partial w^{(4)}} \\ \dfrac{\partial v^{(1)}}{\partial u^{(4)}} & \dfrac{\partial v^{(1)}}{\partial v^{(4)}} & \dfrac{\partial v^{(1)}}{\partial w^{(4)}} \\ 0 & 0 & \dfrac{\partial w^{(1)}}{\partial w^{(4)}} \end{bmatrix}, \qquad (16)$$

where we have again chosen the parametric coordinate such that surface $a$ is a constant level surface of $w^{(1)}$ and surface $c$ is a constant level surface of $w^{(4)}$ such that, on surface $a$ we have

$$\frac{\partial w^{(1)}}{\partial v^{(4)}} = \frac{\partial w^{(1)}}{\partial u^{(4)}} = 0. \qquad (17)$$

Using Eqs. (15) and (16), the relations of the tangential fields on surface $c$ in illusion space and surface $a$ are given by

$$\begin{aligned} E_{u^{(4)}}^{(4)} &= \frac{\partial u^{(1)}}{\partial u^{(4)}} E_{u^{(1)}}^{(1)} + \frac{\partial v^{(1)}}{\partial u^{(4)}} E_{v^{(1)}}^{(1)}, \\ E_{v^{(4)}}^{(4)} &= \frac{\partial u^{(1)}}{\partial v^{(4)}} E_{u^{(1)}}^{(1)} + \frac{\partial v^{(1)}}{\partial v^{(4)}} E_{v^{(1)}}^{(1)}. \end{aligned} \qquad (18)$$

By comparing Eqs. (12) and (18), it is clear that on surface $c$,

$$\begin{aligned} E_{u^{(4)}}^{(4)} &= E_{u^{(3)}}^{(3)}, \\ E_{v^{(4)}}^{(4)} &= E_{v^{(3)}}^{(3)}, \end{aligned} \qquad (19)$$

if and only if on surface $a$:

$$\begin{aligned} \frac{\partial u^{(2)}}{\partial u^{(3)}} &= \frac{\partial u^{(1)}}{\partial u^{(4)}}, \quad \frac{\partial v^{(2)}}{\partial u^{(3)}} = \frac{\partial v^{(1)}}{\partial u^{(4)}}, \\ \frac{\partial u^{(2)}}{\partial v^{(3)}} &= \frac{\partial u^{(1)}}{\partial v^{(4)}}, \quad \frac{\partial v^{(2)}}{\partial v^{(3)}} = \frac{\partial v^{(1)}}{\partial v^{(4)}}. \end{aligned} \qquad (20)$$

Since both **A** and **B** map surface $c$ to $a$, we can always choose **A** and **B** such that they map the same point on surface $c$ to the same point on surface $a$. Accordingly, Eq. (20) can be fulfilled. With that, we have proved that the tangential fields on surface $c$ are the same for both the real space and the illusion space. For the tangential field on surface $d$, since **B** maps each point of surface $d$ back to itself, similar to the case of boundary condition matching on surface b, it can be easily seen that the tangential fields on surface $d$ are exactly the same for both the real space and the illusion space. Since surfaces $c$ and *d* together form a closed surface, and both fields on surfaces *c* and *d* are the same for both the real space and the illusion space, by the uniqueness theorem, the field outside surfaces *c* and *d* for both the real space and the illusion space are exactly the same. With that, we have disguised the object into the illusion and thus completed our proof.

We note that while our proof here is for three-dimensional geometries, it can be easily generalized to two dimensions. Moreover, it can also be generalized to the case in which the illusion device does not share a part of its boundary with the virtual boundary, i.e. surface *d*, as Fig. A3 shows. In this case, the restoring medium is completely surrounded by the complementary medium.

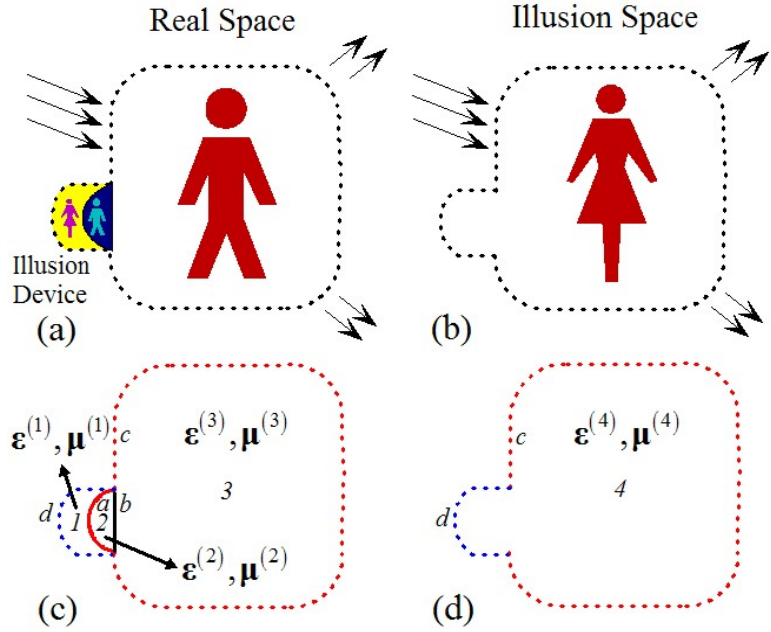

Fig. A1. The working principle of an illusion device that transforms the stereoscopic image of the object (a man) into that of the illusion (a woman). (a) The man (the object) and the illusion device in real space. (b) The woman (the illusion) in the illusion space. (c) The physical description of the system in real space. The illusion device is composed of two parts, the complementary medium (region 2) that optically "cancels" a piece of space including the man (region 3), and the restoring medium (region 1) that restores a piece of the illusion space including the illusion (region 4 in (d)). Both real and illusion spaces share the same virtual boundary (dashed curves).

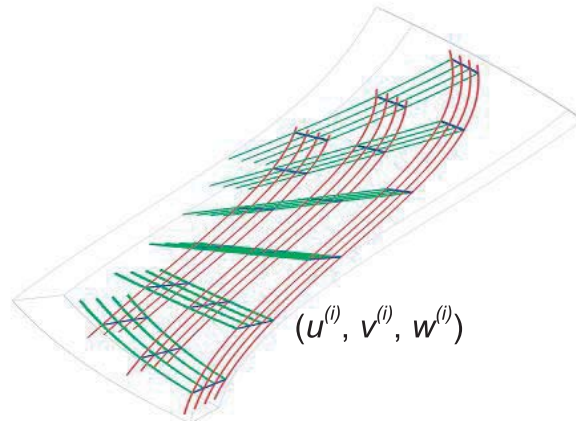

Fig. A2. An illustration of an arbitrary curved coordinate system.

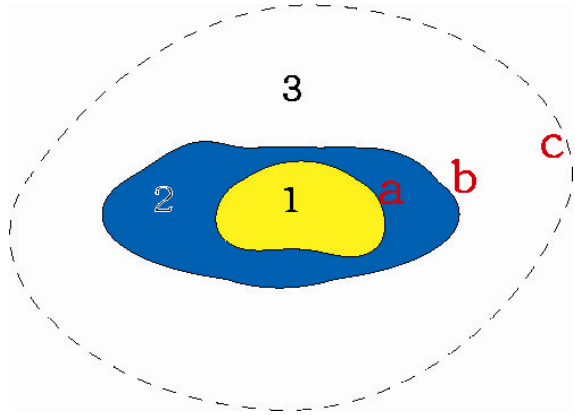

Fig. A3. Another topology of illusion device, in which the restoring medium (region 1) is completely surrounded by the complementary medium (region 2). The boundary of region 3 (curve *c*) is the virtual boundary.

Part B: Numerical demonstration of the illusion optics by using the system in Fig. 2(b) under various kinds of incident waves to show that the device functionality is independent of the form of the incident waves.

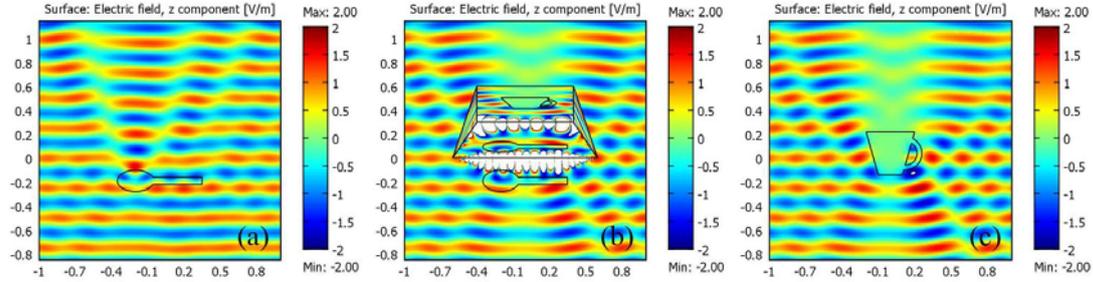

Fig. B1. A TE plane wave of wavelength 0.25 unit incident from below.

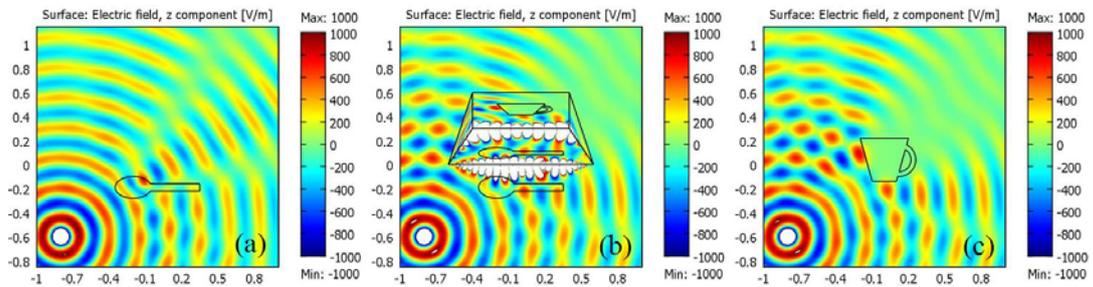

Fig. B2. A TE point source of wavelength 0.25 unit placed at (-0.8, -0.6).

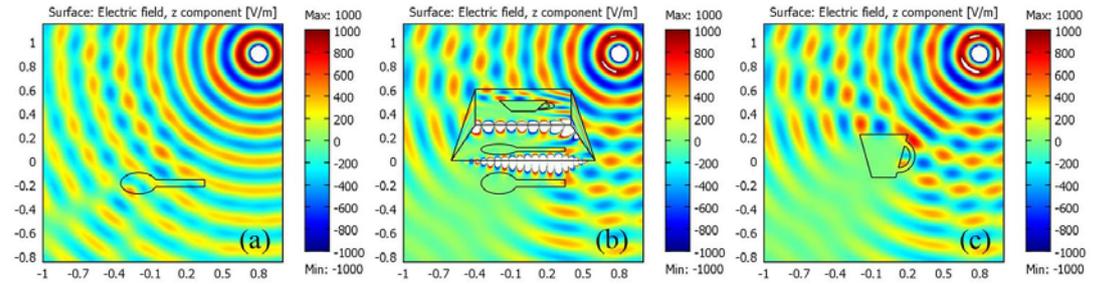

Fig. B3. A TE point source of wavelength 0.25 unit placed at (0.8, 0.9).

From these numerical simulation results, it can be clearly seen that the illusion optics effect is independent of the incident angle and profile of the incident waves.

Part C: Description of the illusion device demonstrated in Fig. 3(b), and a numerical simulation of revealing an object hidden inside a container.

The illusion device in Fig. 3(b) is composed of four parts. The left trapezoidal part in contact with the wall is the complementary medium with $\varepsilon_z^{(2)} = 2$, $\mu_x^{(2)} = -0.5$ and $\mu_y^{(2)} = -2$, formed by a coordinate transformation of $x^{(2)} = -x^{(3)}/2$. Here, the complementary medium is only negative in permeability because the "cancelled" wall is negative in permittivity (i.e., metallic). The upper and lower triangular parts and the middle rectangular part on the right constitute the restoring medium. The upper and lower triangular parts are composed of a medium with $\varepsilon_z^{(1)} = 4$, $\mu_{xx}^{(1)} = 9.25$, $\mu_{yy}^{(1)} = 4$ and $\mu_{xy}^{(1)} = \mp 6$, formed by the coordinate transformations of $x^{(1)} \pm 2(y \mp 0.5) = 1/4 \cdot \left( x^{(4)} \pm 2(y \mp 0.5) \right)$, respectively. The middle rectangular part is composed of a medium of $\varepsilon_z^{(1)} = 4$, $\mu_x^{(1)} = 0.25$ and $\mu_y^{(1)} = 4$, formed by the coordinate transformation of $x^{(1)} - 0.2 = 1/4 \cdot \left( x^{(4)} - 0.2 \right)$. Since the aim is to create a piece of free space in this case, there is no compressed version of any illusion object inside the restoring medium. This "super-vision" illusion device does not require a broad bandwidth and thus can be constructed by resonant metamaterials designed at a single selected working frequency.

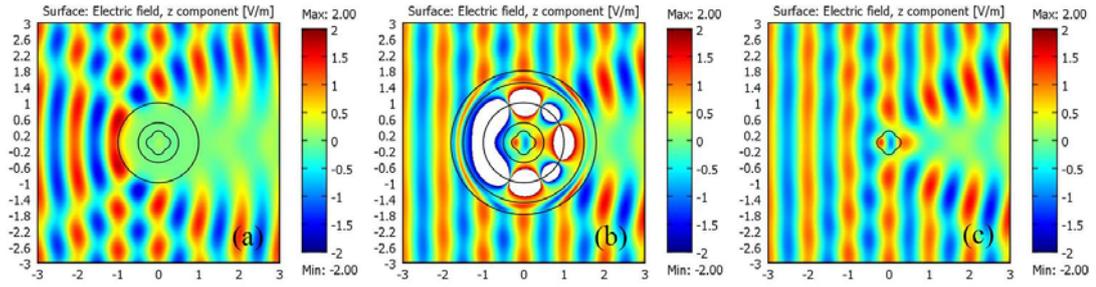

Fig. C1. A numerical demonstration of revealing an object hidden inside a container by using illusion optics. (a) An object of $\varepsilon = 5$ is hidden inside a circular shell of $\varepsilon = -1$ (metallic), such that a TE plane wave incident from the left cannot "see" the object. (b) A circular illusion device consisting of an inner circular layer of complementary medium that optically "cancels" the shell, and a circular layer of restoring medium that restores a circular layer of free space, is placed outside the shell. It is clearly seen that the scattering pattern outside the device is now changed into exactly the same pattern as the scattering pattern of the object itself, as is shown in (c).